**Conserved Linking in Single- and Double-Stranded Polymers**

**Joseph S. Plewa and Thomas A. Witten**

*James Franck Institute and Department of Physics, University of Chicago,*

*5640 S. Ellis Ave., Chicago, IL 60637*

ABSTRACT:   We demonstrate a variant of the Bond Fluctuation lattice Monte Carlo model in which moves through cis conformations are forbidden.  Ring polymers in this model have a conserved quantity that amounts to a topological linking number.  Increased linking number reduces the radius of gyration mildly.  A linking number of order 0.2 per bond leads to an eight-percent reduction of the radius for 128-bond chains.  This percentage appears to rise with increasing chain length, contrary to expectation.  For ring chains evolving without the conservation of linking number, we demonstrate a substantial anti-correlation between the twist and writhe variables whose sum yields the linking number.  We raise the possibility that our observed anti-correlations may have counterparts in the most important practical polymer that conserves linking number, DNA.

**I. Introduction**

One may distinguish two types of flexible, linear polymer chains, as pictured in Figure 1: those that relax to equilibrium after being twisted (like polyethylene), and those that do not (like DNA). These non-relaxing chains obey a conservation law. For each chain configuration one may define a "linking number", and this linking number cannot change with time. The linking number of any closed, double-stranded chain is defined as the number of times one strand must be passed through the other in order to separate the strands. The consequences of conserved linking for DNA and for macroscopic cords are well appreciated.[1-7] A large imposed linking number leads such a chain to twist upon itself. This is the phenomenon of supercoiling. Supercoiling is a way of effecting global change in the properties of a chain by changing local mechanical structure in a small region. As such, it represents a powerful mechanism for controlling the chain---a mechanism that may be important for biological processes. Such far-reaching consequences are to be expected for any chain with conserved linking.

This paper aims to explore the essential consequences of conserved linking by considering a primitive realization. Our realization is a lattice model that embodies conserved linking with a single strand and without the worm-like rigidity of DNA. The model resembles a hydrocarbon chain with a certain local restriction on the rotation of its backbone bonds. This chain proves to have properties quite different from those of DNA. It has strong excluded-volume swelling and thus shows simultaneous effects of swelling and supercoiling. As expected for polymers that conserve linking number, the chain contracts upon twisting. However, this contraction does not scale in the expected way with chain length. Another anomalous feature appears in the partitioning of imposed linking number into "twist" and "writhe". The linking number of any chain may be decomposed into twist, a locally defined quantity, and writhe, a quantity that depends only on the backbone configuration. Our model shows a strong anti-correlation of twist with writhe, which is not generally anticipated in DNA.[8-12] The presence of this correlation in our model raises the possibility of such correlations in DNA.

The statistics of twist and writhe in self-avoiding walks have been much explored in the literature.[13-18] The influence of linking number on DNA has been studied extensively via continuum models .[19-23] Recently a powerful correspondence has been worked out between these continuum models and rigid-body dynamics.[24] Our work is rather in the alternative spirit of the work of Orlandini, Whittington and their collaborators.[13-18] These authors have recognized the value of the lattice polymers as a way to study statistical issues of twist and writhe of interest for DNA. For simple lattice ring polymers, they have demonstrated that the mean-squared writhe scales linearly with the chain length. They have shown that this writhe is positively correlated with torsion. They have invented a lattice ribbon model and have demonstrated that its mean-squared twist also scales linearly with chain length in an ensemble of unconstrained linking number. The lattice model introduced below complements the previous ribbon polymer model. It is quite different from the ribbon model in detail, yet it shows the same macroscopic scaling properties. We've used our model to explore properties not considered previously. In addition, unlike the ribbon model, our model is constructed to conserve linking number. This makes it convenient for studying the dynamic and ergodic consequences of conserved linking number.  Recently, Velichko, Yoshikawa, and Khokhlov[25] have published a study of a link-conserving, double-stranded chain.

We begin by describing the model and our Monte Carlo simulation embodying it.  We report on the performance of the simulation and describe the crosschecks we used to test its validity.  Next we recall the definitions of twist and writhe and report on their statistical behavior in our chain.  Twist, writhe and linking number all have mean-squared averages that grow linearly with chain length, as expected.  In addition our chain shows twist-writhe correlations even when the linking number is allowed to vary freely.  These correlations correctly predict the response to an imposed linking number, via the fluctuation-dissipation theorem.  Next we report how the size and shape of the chain respond to imposed changes in linking number, showing unexpected scaling with chain length.  Then in the discussion section we explore the implications of our results.  We show how a twist-writhe correlation naturally arises in our model from a mechanically-induced coupling of twist with local torsion.

## II. Description of Model and Simulation

We model the polymer chain as a self-avoiding walk on a lattice, in which we allow nearest neighbor, next nearest neighbor, and further local steps as described below.   We include the additional restriction that no two adjacent segments may be collinear.  This restriction is realistic for hydrocarbon polymers, and guarantees that any segment and its successor define a unique plane, which in turn has a non-degenerate normal vector.  The relative orientation of two adjacent normal vectors establishes whether a sequence of three segments is in the so-called cis configuration (Figure 2). All configurations cost zero energy, except for the cis configuration, which is forbidden. Likewise, Monte Carlo moves which traverse a cis configuration are forbidden.

If the ends of such a chain are joined to form a ring, this ring has a linking number that is conserved by the Monte Carlo dynamics. Thus the simulated ring conserves linking number. To see this, we define a partner strand by connecting the tips of the normal vectors for a given ring chain configuration. One may readily verify that rotating a given segment through the cis configuration (Figure 2b) causes the partner strand to cross the chain. If the perpendicular vectors are made arbitrarily short and the chain does not cross itself, then the partner strand crosses the chain if and only if the three segments including the intersected segment are in the cis configuration. By forbidding cis configurations we forbid these intersections. Thus the linking number is conserved provided the chain does not intersect itself.

The Monte Carlo simulation employs the bond-fluctuation algorithm developed by Carmesin and Kremer.[26] This algorithm allows simulation of self-avoiding walk ring polymers, by using a fine-grained lattice, such that step sizes used in randomly moving nodes are smaller than the bond lengths, which are variable. Bond vectors and Monte Carlo moves are chosen so that segments are in allowed configurations, and cannot intersect during a move. Since the segments do not intersect, linking number is conserved, as shown by the above argument.

For each chain configuration, the partner strand is assigned by taking the cross product of the two adjacent vectors at each vertex, and placing the partner node a small, fixed distance along this perpendicular vector. For each successive node, the cross product is reversed. Thus if $\boldsymbol{\tau_i}$ is the vector pointing from the (i-1)th node to the ith node, the displacement vector $\mathbf{u_i}$ between the ith node and its counterpart on the partner strand is defined to be in the direction $(-1)^i \boldsymbol{\tau_i} \times \boldsymbol{\tau_{i+1}}$. Then the partner nodes are connected to form the partner strand, as shown in Figure 2a. For this staircase or "trans" configuration, the partner strand forms a staircase parallel to the chain.

The chain is evolved by randomly choosing a node along the chain, and then selecting a random nearest-neighbor site for that node. The node is moved to the selected site if this move is allowed. The allowed bond vectors are shown in Figure 3. A lookup table streamlines the testing for forbidden moves. Local intersections, collinearity of adjacent segments, invalid bond vectors, and rotations through the cis configuration are all a property of at most three adjacent segments and the candidate move of one of their nodes. The lookup table is constructed by checking every possible move of each node in every possible three-segment sequence. During simulation, moves are checked by simply looking at the entry appropriate for that move and sequence in the table. The only feature of the chain that the simulation must calculate as the chain evolves is non-local intersection of nodes. This is accelerated by use of a memory image of the lattice so that only one check is required for a given candidate move.

An additional benefit of the lookup table method is that tables with different restrictions can be employed. These tables, along with a toggle on the global intersection test, allow simple changes between a variety of simulation rules. In addition to the table describing our link-conserving polymer, we employed tables with no twist constraints, using both random walk statistics and self-avoiding walk statistics. We measured auto-correlation functions for the radius of gyration to determine the relaxation time for link-conserving and non-conserving chains. A link-conserving chain with 64 bonds requires roughly $2 \times 10^6$ attempted moves to attain a statistically independent configuration. Without the link-conserving constraint, the number of attempts decreases by about 25%. A typical self-avoiding walk ring chain which conserves link is shown in Figure 4, with its partner strand.

Our Monte Carlo procedure obeys detailed balance just as the original Carmesin Kremer procedure does. Our procedure is the same as this one except for the exclusion of certain moves: namely, moves through a cis configuration. Since the exclusion is the same in either direction, it does not disturb the detailed balance. Despite this detailed balance our simulation does not explore all self-avoiding configurations with a given linking number. For example it does not explore knotted configurations of the backbone.

We did extensive tests of both random walks and self-avoiding walks to ensure that the simulation behaved correctly. We simulated both open and ring chains. Figure 5 shows the radius of gyration and the root-mean-squared end-to-end length as a function of chain length for an open random walk chain. Figure 6 shows the corresponding quantities for a self-avoiding walk. Similar results were found for ring chains. Figures 7 and 8 show the structure factor versus wave-vector $\mathbf{q}$ for a variety of chain lengths, for random walk and self-avoiding walk chains respectively. The inverse length $\mathbf{q}$ is normalized so that the data collapse onto one curve. In all cases, the scaling was as expected, and consistent with results of other simulations.[27] Again, the correct scaling was also found for ring chains.

## III. Link, Twist, and Writhe Statistics

As with DNA, we can describe our chain and associated partner strand using the linking number Lk, and its decomposition into twist Tw and writhe Wr using White's theorem[31]

$$Lk = Tw + Wr. \tag{1}$$

To define these quantities, we adopt a continuum representation with the chain described by a position vector $\vec{R}(s)$ for each position s along the chain. The partner strand has an analogous position $\vec{R}'(t)$. Linking number denotes the number of times the two strands wrap around each other, and is given by the Gaussian integral

$$Lk \quad = \quad \frac{1}{4\pi} \oint\oint ds\, dt\, \hat{t}(s) \times \hat{t}'(t) \cdot \frac{(\vec{R}(s) - \vec{R}'(t))}{\left|(\vec{R}(s) - \vec{R}'(t))\right|^3} \tag{2}$$

where $\vec{R}(s)$ and $\vec{R}'(t)$ denote the position vectors of the true chain and the partner strand, and the $\hat{t}$'s denote the tangent vectors to the curves. It is easily shown using Stokes' theorem that this formula yields an integer equal to the net number of times the chain and the partner strand cross when viewed in any projection. A crossing is positive if the $\hat{t}$, $\hat{t}'$, and $\vec{R} - \vec{R}'$ at the intersection form a right-handed coordinate system; otherwise it is negative. Likewise it can be shown that if the chain and partner strand can be separated without crossing, Lk = 0.

Tw is given by

$$Tw \quad = \quad \frac{1}{2\pi} \oint ds \, \hat{t}(s) \times \hat{u}(s) \cdot \dot{\hat{u}}(s) \tag{3}$$

where $\hat{u}$ is the unit vector perpendicular to $\hat{t}$ which points from the chain to the partner strand. Twist is the integrated rotation of projection of the vector from the backbone to the partner strand along the plane perpendicular to the backbone at each point, divided by $2\pi$. Writhe is given by

$$Wr \quad = \quad \frac{1}{4\pi} \oint \oint ds \, ds' \, \hat{t}(s) \times \hat{t}(s') \cdot \frac{(\vec{R}(s) - \vec{R}(s'))}{\left| (\vec{R}(s) - \vec{R}(s')) \right|^3} \tag{4}$$

where now the integrals are over the same curve. Writhe can be described as the sum of signed self-intersections of the projection of the chain averaged over all possible views.[32]

In our simulations we calculated the linking number as the sum of the signed intersections of the projection of the two strands onto a plane, as described above. We chose a projection direction incommensurate with the lattice, so that all intersections occur at non-zero angles. The twist was calculated using (3). We calculated the writhe for many configurations as a consistency check using (4). Thus we verified that our calculated linking number, twist and writhe were consistent with White's Theorem. All calculations used the polymer chain as one strand, and the partner strand as the second strand. This contrasts with the convention used for DNA, in which the backbone is defined as an imaginary line running down the center of the helix, and either sugar-phosphate chain is used as the partner strand.[2]

Ring chains of length 16 to 256 beads were simulated using a lookup table that enforced all constraints and excluded cis configurations. Chains with various linking numbers were simulated for thousands of relaxation times, and each preserved its original linking number throughout its run. A lookup table which enforced all constraints but allowed cis configurations was then used to study the statistical fluctuations in linking number. As expected the probability of a particular linking number occurring had a Gaussian distribution (Figure 9), with standard deviations that grew with chain length as $\sigma = 0.23 \, N^{.52}$ (Figure 10).

Deviations in twist were calculated for these chains as well. The mean-squared twist for all the chains is shown is in Figure 11 as a function of N. We find that $\left\langle Tw^2 \right\rangle_0 = .0767 \, N$, where the subscript 0 denotes averages taken in an ensemble where linking number is allowed to fluctuate freely. The average of LkTw was also measured; this allows calculation of the RMS writhe via

$$\left\langle Wr^2 \right\rangle_0 = \left\langle Lk^2 \right\rangle_0 + \left\langle Tw^2 \right\rangle_0 - 2 \left\langle LkTw \right\rangle_0 . \tag{5}$$

We find that $\left\langle Wr^2 \right\rangle_0 = .0231 \, N$ as shown in Figure 11. We note that our scaling for the RMS twist, writhe, and linking number is consistent with that of the ribbon model of Orlandini and Whittington.[13-18] Finally, we calculated the twist-writhe correlation using

$$\langle TwWr \rangle_0 = \langle LkTw \rangle_0 - \langle Tw^2 \rangle_0 . \qquad (6)$$

The observed twist-writhe correlation (Figure 11) is a substantial fraction of $\langle Tw^2 \rangle_0$ and $\langle Wr^2 \rangle_0$; in DNA, such correlations are normally neglected.[8-12]

We artificially increased the linking number for chains with N = 16 to 256 segments to values up to $N^{-5}$ in integral steps in order to study the response of the chains to imposed Lk. By constraining the imposed linking number, it is possible to measure how much of an imposed Lk goes into twist and how much into writhe. Using the fluctuation-dissipation theorem, $\langle Tw \rangle / Lk$ and $\langle Wr \rangle / Lk$ can be calculated as in Marko.[1] Including twist-writhe coupling yields

$$\frac{\langle Tw \rangle}{Lk} = \frac{\langle Tw^2 \rangle_0 + \langle TwWr \rangle_0}{\langle Tw^2 \rangle_0 + \langle Wr^2 \rangle_0 + 2\langle TwWr \rangle_0} , \qquad (7)$$

and

$$\frac{\langle Wr \rangle}{Lk} = \frac{\langle Wr^2 \rangle_0 + \langle TwWr \rangle_0}{\langle Tw^2 \rangle_0 + \langle Wr^2 \rangle_0 + 2\langle TwWr \rangle_0} . \qquad (8)$$

Substituting the averages from the simulation with linking number unconstrained gives $\langle Tw \rangle / Lk = 0.93$ and $\langle Wr \rangle / Lk = 0.07$. The results from the simulation with linking number conserved converge nicely to this prediction. Already at 256 beads $\langle Tw \rangle / Lk =$ 0.94 and $\langle Wr \rangle / Lk = 0.06$. Omitting $\langle TwWr \rangle$ from the analysis results in significant discrepancy with the simulation, $\langle Tw \rangle / Lk = 0.77$ and $\langle Wr \rangle / Lk = 0.23$.

## IV. Dependence of Gyration Radius on Chain Length and Imposed Linking Number

The radius of gyration as a function of linking number for a chain of length N = 128 is shown in Figure 12. Linking numbers greater than 10 were reached by manually winding the chain to the desired Lk. For all chains studied from N = 32 to 256, the gyration radius decreases parabolically with linking number, $R_g = R_{g0}(1 - a(Lk)^2)$. The radius of gyration at Lk = 0 scales robustly as $N^{.6}$, as in Figure 6. The coefficient $a$ as a function of N is plotted in Figure 13. The scaling here is between $N^{-1.25}$ and $N^{-1.5}$. Even linking numbers significantly greater than those encountered in equilibrium did not alter the size of the chain significantly.

It is typically claimed[1] that increasing the linking number by about one unit per persistence length significantly alters the equilibrium chain configurations and thus should alter $R_g$ by an appreciable factor. This criterion corresponds to $a \sim N^{-2}$. This scaling is not inconsistent with the data.

Seeing no gross distortions, we decided to investigate the effect of the imposed linking number on the aspect ratio of the chain. The moment of inertia matrix was calculated and diagonalized. The principal moments were then sorted and averaged. Typical ratios were approximately 2.57:2.07:1. Changing the linking number over the range $\pm N^{-5}$ resulted in no significant change in this ratio.

The coupling of twist and writhe accounts for these observations to some extent. Because of the coupling, most of the imposed link goes into twist. Since writhe describes contortion of the chain, a necessary condition for significant distortion beyond that present in thermal fluctuations is that $\langle Wr \rangle > \langle Wr^2 \rangle^{0.5}$. None of the chains simulated realized this condition. Further, the observed values of writhe were mostly smaller than 1. Even manipulating the chain to have a high imposed linking number (Lk = 20 for N = 128 beads) barely reached this regime because there is a maximum attainable linking number set by the minimum number of segments required for each unit of imposed link. However, as longer chains are examined, this highly writhed regime can be explored.

**V. Discussion**

The most striking result of our simple model is the strong correlation between twist and writhe. Writhe and twist are typically assumed to be independent degrees of freedom, correlated only when a linking number constraint is imposed. This statistical independence is often exploited for the sake of efficiency in computation by simulating only the backbone of the chain, giving information about the non-local quantity writhe. Twist is then introduced analytically using White's theorem and the assumption of independence with writhe. Clearly this approach would not be applicable to our lattice chains.

Our twist-writhe correlation can be understood naturally by considering 3-bond segments of our chain. This is the minimum length segment for which both twist and writhe are defined. Considering a planar, "staircase" segment, it is apparent that both twist and writhe equal zero. Introducing twist by rotating an end segment necessarily makes the configuration non-planar, inducing writhe. Twists of the end segment create helices of the opposite sense, and so introduce writhe of the opposite sign (Figure 14). The correlation is an inevitable consequence of the identification of the binormal of our backbone curve with the twist vector from the backbone to the partner strand. We suggest that energetic constraints in DNA could result in a similar correlation between the binormal and twist vector and result in a twist-writhe correlation.

## VI. Conclusion

    This work demonstrates two novel aspects of constrained linking number in a ring polymer. First, we showed that this constraint is meaningful in the context of single-backbone chains such as polyethylene. We showed that any chain in which cis configurations are forbidden must have a conserved linking number. In real hydrocarbons such as polyethylene, cis configurations are highly unfavorable energetically, resulting in strong suppression of the rate of crossing the cis barrier.[33] For hydrocarbons with more bulky side-chains, the suppression can only be stronger. For example, the barrier to rotation in polystyrene has been calculated to be roughly 10 kT,[34] corresponding to a nominal rotation time in the range of microseconds.[35] Bulkier side groups could increase this barrier height substantially. The rotation time might then exceed, for example, the stress relaxation time. Such conserved-linking effects had not

previously been contemplated in single-backbone polymers. We are investigating the quantitative consequences for hydrocarbon rings and open chains.

A further novel aspect is our finding of substantial twist-writhe correlation, even in chains where the linking number is free to vary. The microscopic origin of this correlation in our lattice chains is readily apparent. The occurrence of these correlations in this simple case leads us to ask what the general conditions for such correlations are. We see no general reason that would rule out such correlations in important link-conserving polymers like DNA, even though current models of DNA lack these correlations. We are investigating the minimal additions to the current models that would allow such correlations to be described. We are also investigating the available measurements of DNA chains to determine how much twist-writhe correlation might be present in practice.

Our simulations of twisted ring chains show surprising conformational properties. These are the first simulations showing full excluded-volume swelling and conserved linking simultaneously, to our knowledge. Our rings appear to shrink in response to twisting more than existing theories have anticipated. This may be due to an interaction between writhe and excluded volume yet to be understood.

**Acknowledgements**

We would like to thank John Marko for countless valuable discussions. One of us (JP) would also like to thank David Grier for his support in this work. This work was supported in part by the National Science Foundation under Award Number DMR

9975533 and in part by its MRSEC program under Award Number DMR 9808595.

[35]We estimated the barrier-crossing time $\tau$ as $\tau_0 e^{U/kT}$ with a barrier energy U of 10 kT and an attempt time $\tau_0$ of $10^{-10}$ seconds, c.f. K.J. Laidler, *Chemical Kinetics,* 3[rd] ed. (McGraw-Hill, New York, 1987).

Figure Captions

**Figure 1.**  Sketch of polymers without and with conserved linking.  Left, a flexible hydrocarbon chain attached to an anchor at each end.  If the upper anchor is twisted, the chain rapidly relaxes to an ensemble of configurations indistinguishable from those of the untwisted state.  Right, a double-stranded chain such as DNA.  When its anchor is twisted, this chain explores an ensemble of configurations that remains different from those of the untwisted chain.

**Figure 2.**  Trans and cis configurations of a three-segment chain.  a)  Trans configuration in the shape of a Z.  b)  Cis configuration in the shape of a U.  For the trans configuration, the imaginary partner strand defined in the text is shown as a dashed line.

**Figure 3.**  Allowed bond vectors in the Carmesin-Kremer bond-length fluctuation model. The full set of allowed bonds is obtained from these by lattice rotation.

**Figure 4**.  Typical configuration of a 256-bond chain, shown here after $10^8$ attempted Monte Carlo moves.  The ribbon consists of the backbone chain and its partner strand.

**Figure 5.**  Log-log plot of radius of gyration $R_g$ (lower set) and root-mean-squared end-to-end distance $R_0$ (upper set) versus number of segments for a random walk chain. Straight lines indicate power-law behavior with exponent $\nu = .5$  The offset between the two lines indicates a ratio $R_0/R_g \sim 2.4$, in agreement with other simulations[27] and the exact[28] asymptotic value of $6^{.5}$.

**Figure 6.** Log-log plot of radius of gyration $R_g$ (lower set) and root-mean-squared end-to-end distance $R_0$ (upper set) versus number of segments for a self-avoiding walk chain. Straight lines indicate power-law behavior with exponent $\nu = .59$  The offset between the two lines indicates a ratio $R_0/R_g \sim 2.5$, in agreement with other simulations[27] and theory[29].

**Figure 7.**  Structure factor $S(q)$ versus $qN^{.5}$ using random walk constraints.  The Debye function[30] expected for large N is shown as a solid line.  The upper data set is for N = 32 segments, the lower set is for N = 128 segments.

**Figure 8.** Structure factor $S(q)$[29] versus $qN^{.588}$ using self-avoiding walk constraints.  The solid line shows the expected $q^{-1/.588}$ dependence.  The dashed line shows the corresponding $q^{-2}$ dependence of random walk polymers.  The upper data set is for N = 32 segments, the lower set is for N = 128 segments.

**Figure 9.**  Probability of occurrence of a particular linking number for a chain with 128 segments, indicating the Gaussian form of the probability distribution.

**Figure 10.**  Standard deviation of the probability distribution of Lk versus square root of chain length.

**Figure 11.** Mean-squared Twist (top data set), Mean-squared Writhe (middle), and average of Twist times Writhe (bottom) as a function of chain length, for chains in which the linking number is unconstrained.

**Figure 12.** Radius of gyration as a function of linking number for a 128 bead chain.

**Figure 13.** Coefficient *a* in the parabolic fit of the radius of gyration decrease as a function of imposed Linking Number, multiplied by $N^2$, plotted versus chain length N. According to conventional models,[1] the plotted values should attain a finite asymptote of order unity for large N.

**Figure 14.** Correlation between twist and writhe resulting from rotation creating twist of one sign and a helix with writhe of opposite sign. Top: staircase configuration with no twist or writhe. Middle: Result of imposed clockwise twist. Bottom: Same configuration as middle panel, viewed along the helical axis of the backbone. This backbone is now a counterclockwise helix, which contributes to counterclockwise writhe.

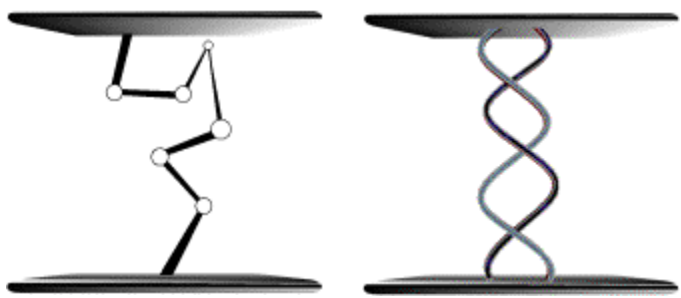

Figure 1.

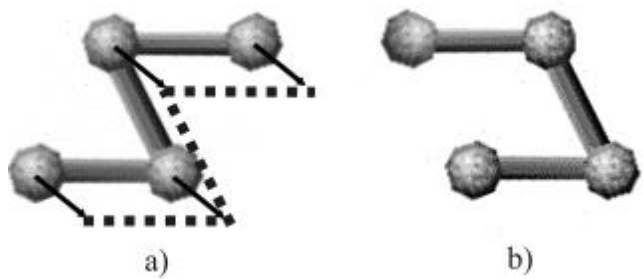

a)                    b)

Figure 2.

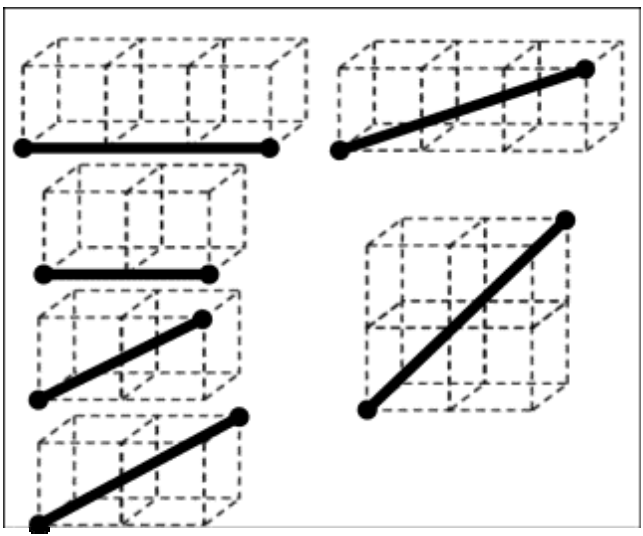

Figure 3.

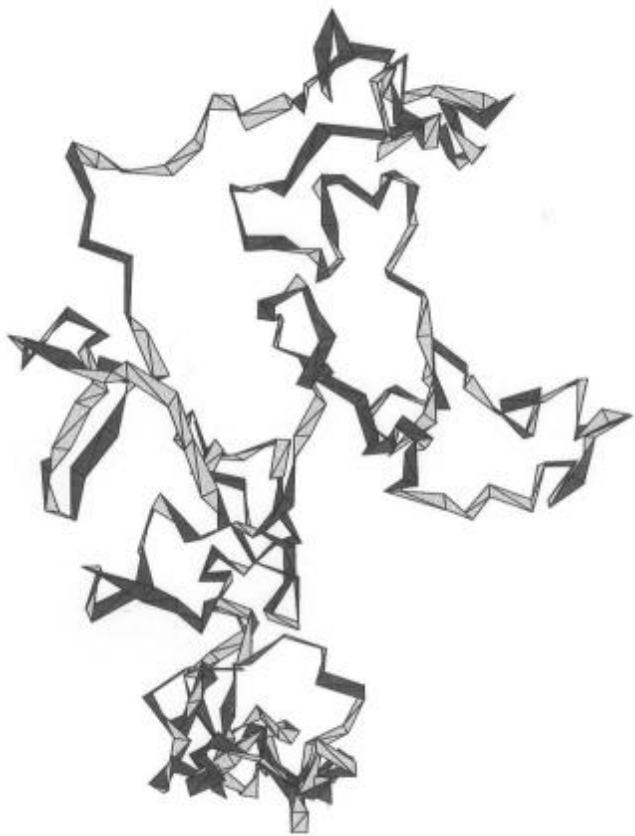

Figure 4.

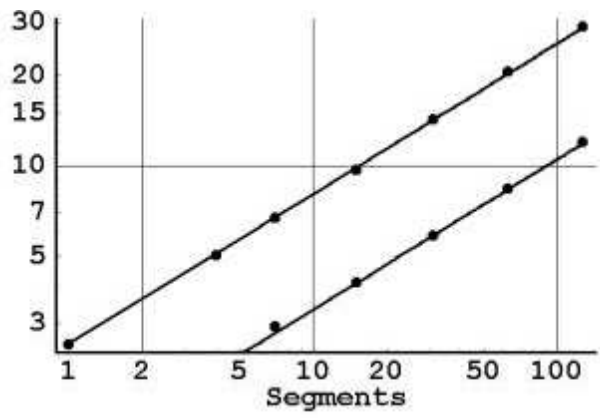

Figure 5.

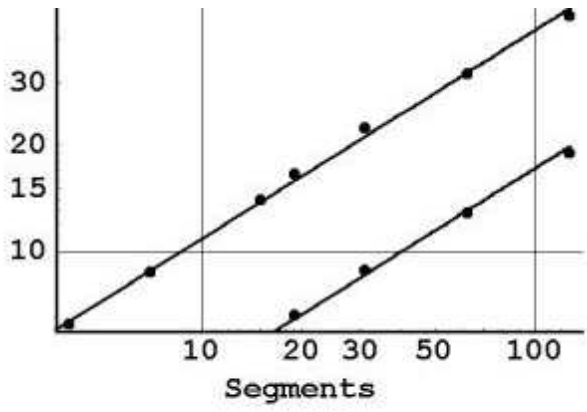

Figure 6.

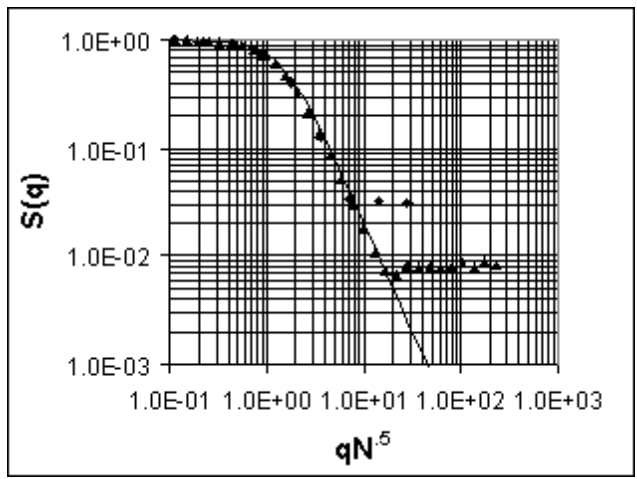

Figure 7.

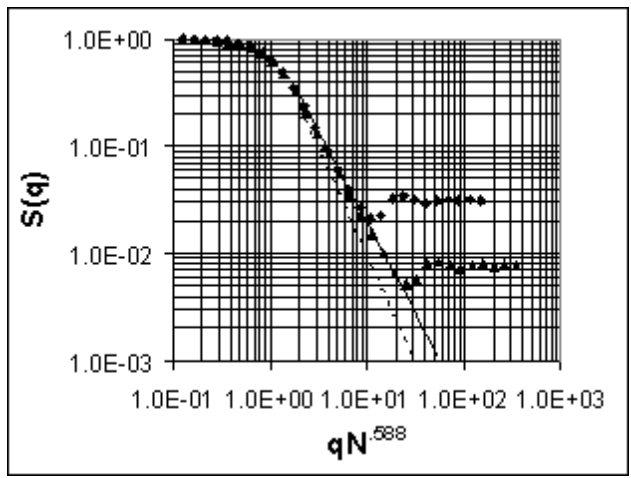

Figure 8.

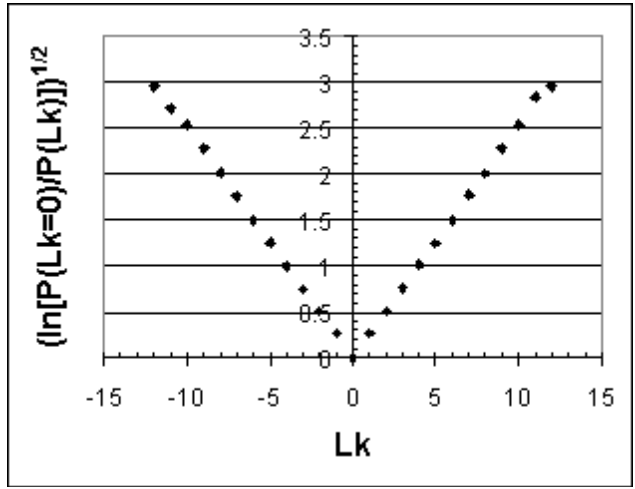

Figure 9.

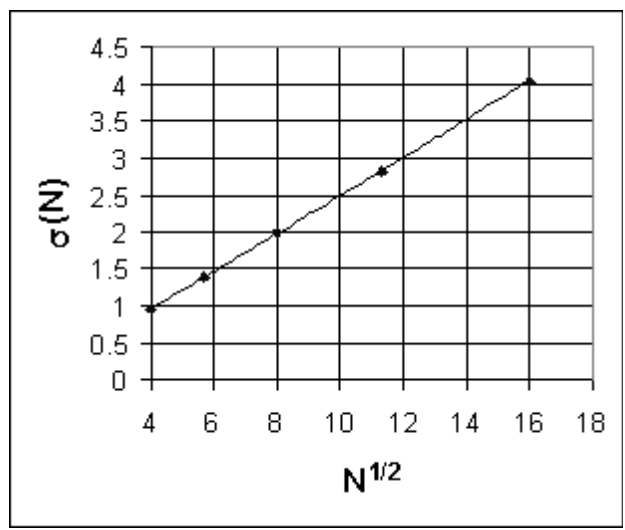

Figure 10.

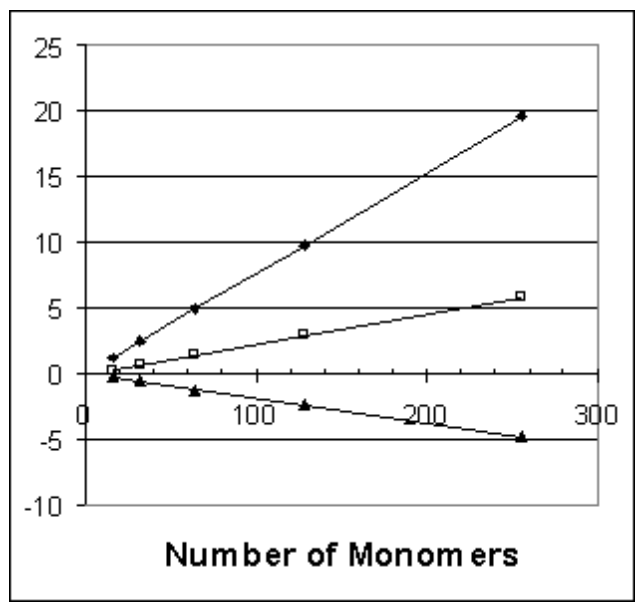

Figure 11.

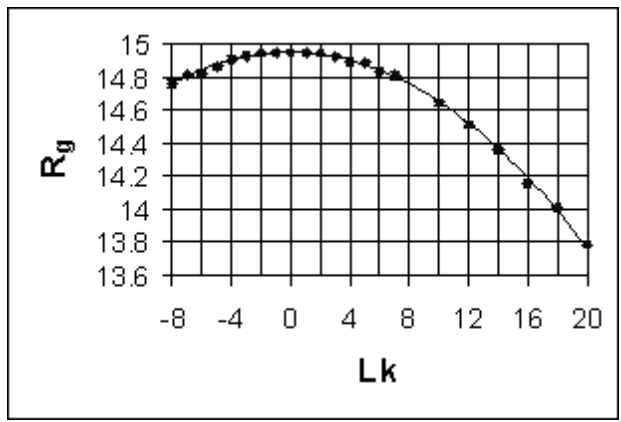

Figure 12.

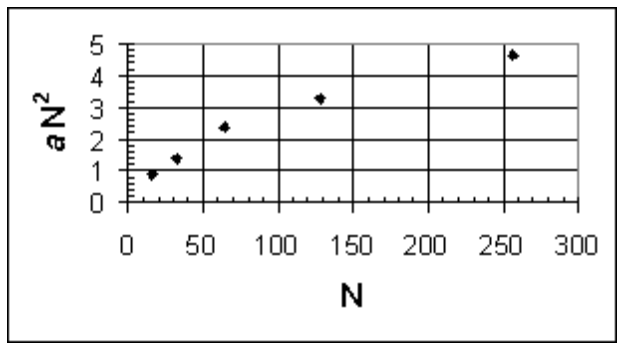

Figure 13.

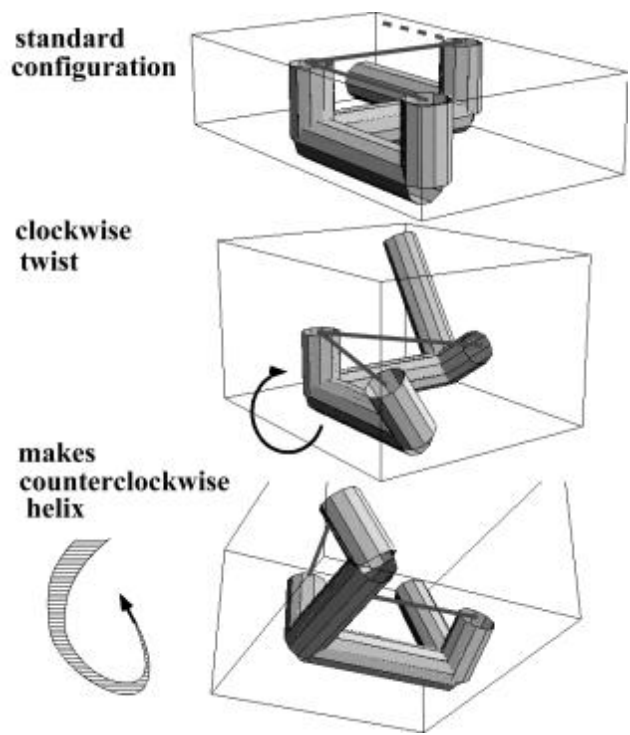

Figure 14.